\documentclass[aps,showpacs,manuscript,12pt]{revtex4}
\usepackage{amssymb}
\usepackage{amsmath}
\usepackage{graphicx}

\begin{document}
\title{\textbf{Generalized covariant gyrokinetic dynamics of magnetoplasmas$^{\S }$}}
\author{C. Cremaschini$^{a}$, M. Tessarotto$^{b,c}$, P. Nicolini$^{b}$ and
A. Beklemishev$^{d}$} \affiliation{\  $^{a}$Department of
Astronomy, University of Trieste, Trieste,Italy, $^{b}$Department
of Mathematics and Informatics, University of Trieste, Trieste,
Italy,  $^{c}$Consortium of Magneto-fluid-dynamics, University of
Trieste, Trieste, Italy, $^{d}$Budker Institute of Nuclear
Physics, Novosibirsk, Russia}
\begin{abstract}
 A basic prerequisite for
the investigation of relativistic astrophysical magnetoplasmas,
occurring typically in the vicinity of massive stellar objects
(black holes, neutron stars, active galactic nuclei, etc.), is the
accurate description of single-particle covariant dynamics, based
on gyrokinetic theory (Beklemishev et al.,1999-2005). Provided
radiation-reaction effects are negligible, this is usually based
on the assumption that both the space-time metric and the EM
fields (in particular the magnetic field) are suitably prescribed
and are considered independent of single-particle dynamics, while
allowing for the possible presence of gravitational/EM
perturbations driven by plasma collective interactions which may
naturally arise in such systems. The purpose of this work is the
formulation of a generalized gyrokinetic theory based on the
synchronous variational principle recently pointed out (Tessarotto
et al., 2007) which permits to satisfy exactly the physical
realizability condition for the four-velocity. The theory here
developed includes the treatment of nonlinear perturbations
(gravitational and/or EM) characterized locally, i.e., in the rest
frame of a test particle, by short wavelength and high frequency.
Basic feature of the approach is to ensure the validity of the
theory both for large and vanishing parallel electric field. It is
shown that the correct treatment of EM perturbations occurring in
the presence of an intense background magnetic field generally
implies the appearance of appropriate four-velocity corrections,
which are essential for the description of single-particle
gyrokinetic dynamics.
\end{abstract}
\date{\today }
\maketitle



\section{Introduction}

From the mathematical viewpoint, it is well known that the
Hamiltonian dynamics of charged single particles can be expressed,
in principle, in arbitrary hybrid (i.e., non-canonical) state
variables. The search of phase-space transformations yielding
\textquotedblleft simpler\textquotedblright\ equations of motion
has motivated in the past theoretical research in physics and
mathematical physics. Among such transformations, a particularly
significant case is provided by the gyrokinetic transformation,
yielding generally hybrid variables defined in such a way that one
of them, the so-called gyrophase angle, results ignorable
\cite{Littlejohn,Hahm}. This transformation, which is typically
based on a perturbative expansion (the so-called Larmor radius
expansion), can only be introduced when particles move in a
sufficiently strong magnetic field. The gyrokinetic description of
particle dynamics is useful because in these variables the
equations of motion for single charged particles generally exhibit
a significantly lower computational complexity (with respect to
the case of non-gyrokinetic variables). This is due to the fact
that the dependence on the fast gyro-angle formally disappears in
this representation. The same approach, nevertheless, can be
adopted in principle also for the formulation of the kinetic and
fluid descriptions of relativistic plasmas, provided the treatment
of these plasmas can be based on single-particle dynamics, i.e.,
radiation effects (such as, in particular, radiation-reaction
effects) are negligible. For a certain class of problems in
plasma-physics and astrophysics, involving the\ study of
relativistic plasma flows in strong gravitational fields, the
adoption of the gyrokinetic formulation results of basic
importance. This involves the description of plasmas in which the
strength of the magnetic field becomes very large, which are
observed or assumed to exist in accretion disks and related plasma
jets around neutron stars, black holes, and active galactic nuclei
\cite{Mohanti,Frank}. The conventional non-relativistic
gyrokinetic theory (GKT) has been previously generalized to
include only some relativistic effects, which are essential for
description of confinement of fusion products in thermonuclear
reactors, namely, the particle itself is considered relativistic,
while its drift velocity is not \cite{Pozzo}. Other approaches
have instead investigated the formulation of covariant
relativistic theories applicable in astrophysical problems. A
first attempt
at a formulation of this type was the Ph.D. thesis of Boghosian \cite%
{Boghosian}. In line with previous non-covariant treatments, this
approach assumed however a flat space-time,\ a vanishing parallel
component of the electric field, ignoring non-linear EM
perturbations and invoking the usual gyrokinetic ordering for
admissible wave forms. This implies that, the Maxwell's equations
in his approach were actually modified to maintain the ordering
properties invoked for the linear EM perturbations. These
deficiencies were corrected by Beklemishev \textit{et al.} (1999-2005, \cite%
{Beklemishev1999,Beklemishev2004,Nicolini2005}) who developed a
fully covariant theory (derivation of the gyrokinetic
transformation) holding in principle for arbitrary curved
space-times. This theory, in particular, is applicable in the
presence of non-uniform and non-linear electromagnetic (EM) and
gravitational fields, as well as arbitrary wave-fields, i.e.,
non-linear perturbations of the EM and gravitational fields. Basic
features of their approach, consistent with the allowance of
relativistic drifts typical of many astrophysical plasmas, were
the assumption of non-vanishing parallel electric field and the
description of particle dynamics in the field-related basis
tetrad, allowing a simpler formulation of the relativistic
equations of motion. In addition, the perturbative theory was
employed to derive the relativistic Vlasov-Maxwell equations
expressed in gyrokinetic variables accurate through second order
in $\varepsilon $ (which is the ratio of the gyro-radius/gyro-time
to the equilibrium gradient length/time), while the wave fields
are taken into account in the second order in the amplitude.
Another remarkable feature of the formulation was to extend the
customary applicability limits of GKT to include wave fields which
can be fast oscillating in time ($\omega \rho _{\mathrm{L}}/c\sim
1$) as well as in space ($k\rho _{\mathrm{L}}\sim 1$), without
involving the
customary assumption $k_{\Vert }\rho _{\mathrm{L}}\ll 1$($k_{\Vert }$%
denoting the parallel wave-vector of the perturbation with respect
to the local direction of the magnetic field). Nevertheless, the
approach of Beklemishev \textit{et al}. contained a limitation
linked to the case of a small or vanishing background electric
field. If $E$ and $B$ denote the
invariant eigenvalues of the electromagnetic (EM) field tensor $F_{\mu \nu }$%
, this happens if there results locally $E/B\ll 1$( denoted as
weak electric field assumption (WEA); see also below) or even it
occurs $E=0$. \ For a certain class of problems in plasma-physics
and astrophysics this limitation may be serious. In particular,
this may occur (in relativistic plasmas) either if:

\begin{enumerate}
\item The plasma is characterized by a strong turbulence. Turbulence can
manifest itself by the possible simultaneous presence of
perturbations characterizing both the EM field and the plasma (to
be described either in terms of kinetic or fluid descriptions). In
such cases, due to strong perturbations, the electric field may
locally vanish or become much smaller than the magnetic field.

\item However, difficulties with the perturbative gyrokinetic treatment
adopted in \cite{Beklemishev1999,Beklemishev2004,Nicolini2005} may
occur for slowly varying (or "quasi-equilibrium") plasmas and even
in the absence of wave fields if WEA occurs. This may happen when
the standard gyrokinetic perturbative theory is carried out at
higher orders in the Larmor-radius expansion.

\item The magnetic field is produced primarily by strong diamagnetic
currents. In fact, even in the absence of an electric field (i.e., letting $%
E=0$), it is well known that electric currents are produced in a
magnetized plasma by gradients of the relevant fluid fields (in
particular, fluid velocity, number density, temperature, magnetic
field intensity, etc.).
\end{enumerate}

The physical motivations of the present investigation are related
to the occurrence of strongly magnetized relativistic plasmas in
curved space-time, which may be strongly characterized by the
presence of EM and kinetic turbulence. In the astrophysical
context they are typically related to the existence of accretion
disks, plasma inflows and outflows and relativistic jets, all
occurring typically close to massive objects, such as neutron
stars, black holes (BH) and active galactic nuclei (Mohanty
\cite{Mohanti}). These plasmas are believed to be dominated by
collisionless, fully relativistic dynamics in which the charged
particles of the plasma interact mutually primarily by means for
the very EM mean field which they produce. This limitation was
overcome by the GKT approach developed by Tessarotto \textit{et
al.} (Ref.1 \cite{Cremaschini2006}) extending the earlier
formulation\textit{\
}\cite{Beklemishev1999,Beklemishev2004,Nicolini2005}. The new
approach is based on key features: 1) the adoption of a
synchronous hybrid Hamilton variational principle to describe
single-particle covariant\ dynamics; 2) the introduction of an
extended gyrokinetic transformation, which includes now also a
suitable 4-velocity transformation and is taken of the form
\begin{equation}
\left( r^{\alpha },u^{\beta }\right) \leftrightarrow
\mathbf{y}^{i}\equiv \left( r^{\prime \alpha },u^{\prime \alpha
}\right) , \label{gyrokinetic transformation}
\end{equation}%
where $r^{\alpha }$ and $u^{\beta }$ coincide with the
countervariant components of the 4-vectors $\mathbf{r}$
(4-position) and $\mathbf{u}$ (4-velocity), $r^{\prime \alpha
}$and $u^{\prime \alpha }$denote the gyrokinetic variables; 3) the
4-velocity transformation is defined in such a way to warrant the
fulfillment of the physical realizability condition for the
4-velocity, $u_{\beta }u^{\beta }=1$, also by the transformed
4-velocity, namely $u_{\beta }^{\prime }u^{\prime \beta }=1;$ \ 4)
the gyrokinetic transformation (\ref{funzionale velcanc}) is
defined in terms of
a power series expansions of the form%
\begin{eqnarray}
r^{\mu } &=&r^{\prime \mu }+\sum\limits_{s=1}^{\infty }\varepsilon
^{s}r_{s}^{\mu }\equiv r^{\prime \mu }+\varepsilon \widehat{r}_{1}^{\mu }(%
\mathbf{y}^{\prime },\varepsilon ),  \label{I} \\
u^{\mu } &=&u^{\prime \mu }\oplus \sum\limits_{s=1}^{\infty
}\varepsilon ^{s}v_{s}^{\mu }(\mathbf{y})\equiv u^{\prime \mu
}\oplus \varepsilon \widehat{v}_{1}^{\mu }(\mathbf{y}^{\prime
},\varepsilon )  \label{II}
\end{eqnarray}%
(\textit{extended gyrokinetic transformation}), where $r_{i}^{\mu
}\left( y^{i}\right) $ and $v_{i}^{\alpha }\left( y^{i}\right) $
($i=1,$..) denote suitable perturbations and the operator $\oplus
$ is defined by the 4-velocity addition law \cite{Pozzo} which
permits to satisfy the physical realizability conditions for all
4-vectors involved ($u^{\mu },u^{\prime \mu }$ and $\varepsilon
\widehat{v}_{1}^{\mu }(\mathbf{y}^{\prime },\varepsilon ) $). This
requires that, if $a_{\mu }$ and $b_{\mu }$ denote two arbitrary
4-velocities, i.e., two unit time-like 4-vectors, such that
$a_{\mu }a^{\mu }=1$ and $b_{\mu }b^{\mu }=1,$ the $\oplus $
operator is defined so that
\begin{equation}
a_{\mu }\oplus b_{\mu }=\frac{a_{\mu }+b_{\mu }}{\sqrt{2(1+a_{\mu }b^{\mu })}%
}  \label{somma 4vel-1}
\end{equation}%
defines a 4-velocity too. In addition, here the perturbations
$r_{i}^{\mu }\left( y^{i}\right) $ and $v_{i}^{\alpha }\left(
y^{i}\right) $ are defined in such a way that both have - by
assumption - a pure oscillatory with respect to the gyrophase
$\phi ^{\prime }.$ In Ref.1 this approach was adopted to treat
finite Larmor-radius effects on particle dynamics in the presence
of non-uniform electromagnetic (EM) and gravitational fields.
Nevertheless, an important issue concerns the extension of the
theory (developed in Ref.1) to take into account also the possible
presence of wave-fields, in principle both EM and gravitational. \
In this paper we intend to extend the applicability range of the
GKT of Ref.1 to such a case. On the other hand since the theory
relies on a perturbative expansion to determine relevant equations
of motion, critical issues are its asymptotic convergence and
accuracy. In this paper we intend to develop a perturbative
solution scheme for the covariant gyrokinetic equations of motion
for a particle immersed in a strong background magnetic field,
which exhibits the correct asymptotic convergence properties even
in the presence of EM perturbations. In detail, the main goals of
the investigation are as follows:

\begin{enumerate}
\item generalization of the approach developed in Ref.1 to include the
treatment of wave-fields (EM and gravitational) perturbations
based on the introduction of a suitable gyrokinetic 4-velocity
transformation;

\item explicit construction of the relevant extended gyrokinetic
transformation appropriate for the gyrokinetic treatment of linear
wave perturbations. In particular, we intend to show that, for
prescribed wave-fields, the 4-velocity transformation can be
reduced to the construction of a unique sequence of Lorentz
transformations to be constructed in such a way to assure a) the
fulfillment of the physical realizability condition $u_{\mu
}u^{\mu }=1$ for the 4-velocity; b) the proper convergence of the
perturbative expansion also in the case of validity of the weak
electric field assumption\textit{.}
\end{enumerate}

These features appear particularly relevant for the treatment of
turbulent plasmas in the astrophysical context, thus providing an
effective tool for use in plasma-physics-related problems in
astrophysics.

\section{Formulation of covariant GKT}

Let us briefly summarize the basic steps of the covariant GKT
approach based on the extended gyrokinetic transformation
(\ref{I}),(\ref{II}). For this purpose, we adopt here a direct
perturbative construction method rather than a non-canonical
Lie-transform method \cite{Remark}. As a starting point, this
involves the introduction of a suitable perturbative expansion
also for the EM $4-$potential $A^{\mu }$ and the metric tensor
$g_{\mu \nu }.$ In particular let us assume that they are of the
form
\begin{eqnarray}
A^{\mu } &=&\frac{1}{\varepsilon }\widehat{A}^{\mu }(r^{\alpha
},\varepsilon
)+\lambda a^{\mu }(r^{\alpha }/\lambda ),  \label{pert-1} \\
g_{\mu v} &=&G_{\mu \nu }(r^{\alpha },\varepsilon )+\lambda g_{\mu
v}^{1}(r^{\alpha }/\lambda ),  \label{pert-2}
\end{eqnarray}%
where $\epsilon $ and $\lambda $ are suitable dimensionless real
parameters, both to be assumed infinitesimal and with $\varepsilon
\ll \lambda $. Here the notation is standard (see Ref.1). Thus,
$A^{\mu }$ and $g_{\mu v}$ are respectively the counter-variant
and covariant components of the EM
4-potential of the metric tensor, while $\frac{1}{\varepsilon }\widehat{A}%
^{\mu }(r^{\alpha },\varepsilon ),$ $G_{\mu \nu }(r^{\alpha
},\varepsilon )$ and $\lambda a^{\mu }(r^{\nu }/\lambda ),\lambda
g_{\mu v}^{1}(r^{\nu }/\lambda )$ denote respectively the
equilibrium terms and the so-called wave-fields, to be identified
with suitable, rapidly-varying perturbations. All functions here
considered are generally assumed to be represented by smooth,
i.e., $C^{(\infty )}$ in functions. In particular, invoking the
gyrokinetic transformation (\ref{I}) for the 4-vector $r^{\alpha
},$ it follows that "equilibrium" fields can be expanded in Taylor
series near the guiding-center 4-position $r^{\prime \alpha }.$
This implies that,
neglecting corrections of order $o(\varepsilon ),$ $\frac{1}{\varepsilon }%
\widehat{A}^{\mu }(r^{\alpha })$ and $G_{\mu \nu }(r^{\alpha
},\varepsilon )$
can be approximated respectively as $\frac{\widehat{A}^{\mu }}{\varepsilon }=%
\frac{\widehat{A}^{\prime \mu }}{\varepsilon }+A_{0}^{\mu }\left[
1+o(\varepsilon )\right] $ and $G_{\mu v}=G_{\mu \nu }^{\prime
}+\varepsilon G_{\mu v}^{1}\left[ 1+o(\varepsilon )\right] .$ Here
the primed quantities are all evaluated, as usual, at the
guiding-center position ($A^{\prime \mu }=A^{\mu }\left( r^{\prime
\alpha }\right) ,$ $G_{\mu \nu }^{\prime }=g_{\mu \nu }\left(
r^{\prime \alpha }\right) ,$ $A_{0}^{\mu }=r_{1}^{\alpha }\partial
_{\alpha }^{\prime }A^{\prime \mu }$ and $G_{\mu v}^{1}\equiv
r_{1}^{\alpha }\partial _{\alpha }^{\prime }G_{\mu v}^{\prime }$),
while the perturbations (of relevant fields) depend generally also
on the gyrophase angle $\phi ^{\prime }$ through the perturbation
$\varepsilon r_{i}^{\alpha }\left( y^{i}\right) .$ The
perturbative scheme converges, at least in an asymptotic sense,
only if one assumes the validity of a suitable ordering conditions
(Larmor radius ordering), implying that EM contributions must
dominate, in a suitable sense, over inertial ones in the
variational functional. Formally this corresponds to replace the
electron electric charge $e$ by $e/\varepsilon ,$ leaving the rest
mass $m_{0}$ unchanged and considering the 4-velocity of order
$o(\varepsilon ^{0})$. To obtain the GKT for particle dynamics the
relevant variational functional $S(\mathbf{y})$
must be expressed in terms of suitable gyrokinetic variables, \ $\mathbf{y}%
^{i},$\thinspace which include in particular the gyrophase $\phi $
describing the fast gyration motion of particles around magnetic
flux lines. \ As shown in Ref.1 \ a convenient formulation can be
achieved by identifying the variational functional with the action
$S(\mathbf{y})\equiv \int_{_{1}}^{2}\gamma (\mathbf{y})+dF,$ where
$\mathbf{y}$ denotes the set of variables $\mathbf{y\equiv }\left(
r^{\mu },u^{\mu },\lambda \right) $
and $\gamma $ is the differential 1-form $\gamma (\mathbf{y})=g_{\mu \nu }%
\left[ qA^{\nu }+u^{\nu }\right] dr^{\mu }+\xi \left[ g_{\mu \nu
}u^{\mu }u^{\nu }-1\right] ,$ with $\xi $ a Lagrange multiplier
and $q=Ze/m_{0}$ the normalized electric charge. It is immediate
to show that phase-space
trajectory of the point particle with 4-position $r^{\nu }$ and 4-velocity $%
u^{\nu }$ results, by construction, an extremal curve determined
by the synchronous variational principle $\delta S(\mathbf{y})=0,$
where the synchronous variations $\delta r^{\mu },\delta u^{\mu
},\delta \xi $ are all considered independent
\cite{Cremaschini2006}. Indeed, the physical realizability
condition $u_{\nu }u^{\nu }=1$ is satisfied only by the extremal
curve. To construct\ the transformations (\ref{I})-(\ref{II})
corresponding to (\ref{pert-1})-(\ref{pert-2}) the perturbations
$r_{i}^{\mu }\left( y^{i}\right) $ and $v_{i}^{\alpha }\left(
y^{i}\right) $ must be suitably determined. As usual, $u^{\prime
\alpha }$ can be expressed in terms of suitable independent
(generally hybrid) gyrokinetic variables, one of which includes by
definition the gyrophase $\phi ^{\prime }.$ This is obtained by
projecting $u^{\prime \alpha }$ along the four independent
directions defining the so-called fundamental tetrad unit 4-vectors $%
e_{a}^{\mu }$ (with $a=0,...,3),$ hereon also denoted $\left(
l^{\prime \mu },l^{\prime \prime \mu },l^{\mu },\tau ^{\mu
}\right) $ [or in symbolic
notation ($\mathbf{l},\mathbf{l}^{\prime }\mathbf{,l}^{\prime \prime }%
\mathbf{,\tau )}$]$,$ where $l^{\prime \mu },l^{\prime \prime \mu
},l^{\mu }$ and $\tau ^{\mu }$ are respectively space-like and
time-like orthogonal unit vectors. In particular, the latter can
always be identified with the EM fundamental tetrad, i.e., the
basis vectors of the EM field tensor, in this case to be evaluated
at the gyrocenter position, $F_{\mu \nu }^{\prime }\equiv F_{\mu
\nu }(x^{\prime \alpha }),$ with eigenvalues $B$ and $E.$ In
particular, the space-like vectors $e_{2}^{\mu },e_{a3}^{\mu })$
are assumed to satisfy the eigenvalue equations $F_{\mu \nu
}^{\prime }e_{2}^{\nu
}=Be_{3\mu },$ $F_{\mu \nu }^{\prime }e_{3}^{\nu }=-Be_{2\mu }.$ Here $%
e_{a\mu }=g_{\mu \nu }^{\prime }e_{a}^{\nu }$ are the covariant
components of $e_{a}^{\mu }$ and the labels $a$ assume a tensor
meaning in the tangent space [so that in terms of the Minkowskian
tensor $\eta =diag\left( 1,-1,-1,-1\right) $ there results $e_{\mu
}^{a}=\eta ^{ab}e_{b\mu }$].

\section{Construction of the extended gyrokinetic transformation}

To illustrate the key new features of the present approach let us
point out
the role of the extended gyrokinetic transformations (\ref{I}) and (\ref{II}%
). GKT is a perturbative theory which can in principle be carried
out at any order - both in $\varepsilon ,$ and $\lambda $ -
provided the perturbations
(i.e., $\varepsilon \widehat{r}_{1}^{\mu }$ and $\varepsilon \widehat{v}%
_{1}^{\mu }$) are suitably small, i.e., in particular there
results
uniformly, in an appropriate subset of phase-space,%
\begin{equation}
\varepsilon \widehat{r}_{1}^{\mu }\sim o(\varepsilon ).
\label{consistency cond}
\end{equation}%
This requirement is clearly necessary for the convergence of the
gyrokinetic perturbative expansion. Hence its must be viewed as a
consistency condition for the validity of GKT. To elucidate the
point, it is sufficient to
consider the perturbative formulation of GKT based on an expansion in $%
\lambda $ which is accurate only to leading order in the
dimensionless parameter $\lambda $. As indicated above, $\lambda $
represents the
magnitude of the wave fields defined by Eqs.(\ref{pert-1}) and (\ref{pert-2}%
). Hence in this case a linear approximation in the wave-fields is
actually considered for the perturbative expansion. \ If the
velocity perturbations in Eq.(\ref{II}) are neglected, one can
prove that to assure that the differential 1-form $\gamma $
results gyro-phase independent, the covariant formulation of GKT
requires that to leading order in $\lambda $ the
following two constraints equations must be satisfied \cite{Beklemishev1999}:%
\begin{equation}
\overset{\sim }{u_{\mu }^{\prime }+}\lambda q\overset{\sim }{a^{\prime }}%
_{\mu }-qr_{1}^{\nu }F_{\mu \nu }^{\prime }=0  \label{172}
\end{equation}%
\begin{equation}
\left\{ \left( u_{\mu }^{\prime }+\lambda qa_{\mu }^{\prime }-\frac{1}{2}%
qr_{1}^{\nu }F_{\mu \nu }^{\prime }\right) \frac{\partial r_{1}^{\mu }}{%
\partial \phi ^{\prime }}\right\} ^{\sim }=\frac{\partial R}{\partial \phi
^{\prime }},  \label{173}
\end{equation}%
where $R$ represents in principle an arbitrary gauge function
(which can always be set $R\equiv 0$). Assuming that $EB\neq 0$
and ignoring possible 4-velocity corrections [see Eq.(\ref{II})],
Eq.(\ref{172}) delivers a formal solution for the perturbation
$r_{1}^{\nu }$ (Larmor-radius 4-vector) which
to leading-order in $\lambda $ yields $r_{1}^{\nu }=\frac{w^{\prime }}{q}%
D^{\prime \nu \mu }\left[ l_{\mu }^{\prime }\cos \phi ^{\prime
}+l_{\mu
}^{\prime \prime }\sin \phi ^{\prime }\right] +\lambda D^{\prime \nu \mu }%
\overset{\sim }{a}_{\mu }.$ Here $D^{\prime \nu \mu }$ is the
inverse Faraday tensor which reads $D^{\prime \mu \nu }\equiv
F_{\mu \nu }^{\prime
-1}=\frac{1}{B}(l^{\nu }\tau ^{\mu }-l^{\mu }\tau ^{\nu })-\frac{1}{E}%
(l^{\prime \nu }l^{\prime \prime \mu }-l^{\prime \mu }l^{\prime
\prime }{}^{\nu }).$ This expression manifestly diverges in the
limit $E\rightarrow 0.$ In particular, for $\left\vert
E\right\vert $ sufficiently small it is obvious that the second
term on the r.h.s. can become arbitrarily large thus violating the
ordering condition (\ref{consistency cond}) (\textit{weak electric
field assumption}). To assure the proper convergence of GKT also
in this case it is manifest that an extended formulation of GKT is
required. In particular, as indicated in Ref.1, it is actually
necessary to introduce a perturbative expansion also for the
4-velocity. This can be conveniently represented in terms of the
4-velocity perturbative expansion (\ref{II}) and by imposing the
4-velocity addition law (\ref{somma 4vel-1}). In such a case the
constraints (\ref{172}) and (\ref{173}) are replaced by the three
equations:
\begin{equation}
\left( u_{\mu }^{\prime }\oplus \lambda v_{1\mu }^{\prime }\right)
^{\sim }+\lambda qa_{\mu }^{\prime }-qr_{1}^{\nu }F_{\mu \nu }=0,
\end{equation}%
\begin{equation}
\left\{ \left( \left( u_{\mu }^{\prime }\oplus \lambda v_{1\mu
}^{\prime }\right) +\lambda qa_{\mu }^{\prime
}-\frac{1}{2}qr_{1}^{\nu }F_{\mu \nu
}^{\prime }\right) \frac{\partial r_{1}^{\mu }}{\partial \phi ^{\prime }}%
\right\} ^{\sim }=\frac{\partial R}{\partial \phi ^{\prime }},
\end{equation}%
\begin{equation}
\left\{ \left[ u_{0}^{\prime }\oplus \lambda u_{01}^{\prime }\right] ^{2}-%
\left[ u_{\parallel }^{\prime }\oplus \lambda u_{\parallel 1}^{\prime }%
\right] ^{2}-\left[ w^{\prime }\oplus \lambda w_{1}^{\prime
}\right] ^{2}-1\right\} ^{\sim }=0.
\end{equation}%
Its is immediate to prove that - to leading-order in $\lambda $ -
these equations can be satisfied identically for arbitrary values
of $E,$ by suitable definition of the 4-velocity perturbation
$\lambda v_{1\mu },$ defined in such away to realize appropriate
\textit{velocity cancellations} for the wave-field perturbations
$\lambda qa_{\mu }$ occurring in the previous equations. In
particular it can be proven that this result can be achieved by
means of the following two Lorentz transformations:

\begin{itemize}
\item \textit{First transformation (Lorentz boost, space-time rotation)}:
Let us first introduce a space-time 4-rotation acting along the directions $%
\mathbf{(\tau ,l)}$ and let us impose that the spatial component
of the transformed velocity cancels identically the corresponding
component of the
wave perturbation $\overset{\sim }{a}$. The 4-rotation takes the form:%
\begin{equation}
\left\{
\begin{array}{c}
\overset{\wedge }{u_{\parallel }^{\prime }}=(u_{\parallel
}^{\prime }\oplus \lambda v_{1\parallel }^{\prime })\cosh \alpha
+(u_{0}^{\prime }\oplus
\lambda v_{10}^{\prime })\sinh \alpha \\
\overset{\wedge }{u_{0}^{\prime }}=(u_{\parallel }^{\prime }\oplus
\lambda v_{1\parallel }^{\prime })\sinh \alpha +(u_{0}^{\prime
}\oplus \lambda
v_{10}^{\prime })\cosh \alpha%
\end{array}%
\right.  \label{boost1}
\end{equation}%
where $\overset{\wedge }{u_{\parallel }^{\prime }}$ e $\overset{\wedge }{%
u_{0}^{\prime }}$ represent the two surviving components of the
4-velocity after the transformation.
\end{itemize}

\begin{itemize}
\item \textit{Second transformation (Lorentz-boost, space-time rotation):}
Let us introduce now a second space-time rotation acting along the
two
remaining directions $\mathbf{(l}^{\prime }\mathbf{,l}^{\prime \prime }%
\mathbf{,\tau )}$, and which leave unchanged the parallel component $\overset%
{\wedge }{u_{\parallel }^{\prime }}$ (along the unit 4-vector $\mathbf{\tau }%
)$ already determined by the first Lorentz boost. We therefore
define the
second 4-rotation:%
\begin{equation}
\left\{
\begin{array}{c}
\overset{\frown }{w}^{\prime }=\left( w^{\prime }\oplus \lambda
w_{1}^{\prime }(\mathbf{y})\right) \cosh \beta +\overset{\wedge }{%
u_{0}^{\prime }}\sinh \beta \\
\overset{\frown }{u_{0}^{\prime }}=\overset{\wedge }{u_{0}^{\prime
}}\cosh \beta +\left( w^{\prime }\oplus \lambda w_{1}^{\prime
}(\mathbf{y})\right)
\sinh \beta%
\end{array}%
\right.  \label{boost2}
\end{equation}
\end{itemize}

After the second transformation the 4-velocity takes the (desired) final form%
\begin{equation}
u_{\mu }=\overset{\frown }{w^{\prime }}\left[ l_{\mu }^{\prime
}\cos (\phi
^{\prime })+l_{\mu }^{\prime \prime }\sin (\phi ^{\prime })\right] +\overset{%
\wedge }{u_{\parallel }^{\prime }}l_{\mu }+\overset{\frown }{u_{0}^{\prime }}%
\tau _{\mu }.  \label{velfinale}
\end{equation}%
As a result one finds that the correct definition of the Larmor
4-vector is
\begin{eqnarray}
r_{1}^{\nu } &=&\frac{1}{q}D^{\prime \nu \mu }\left( u_{\mu
}\right) ^{\sim }+\lambda D^{\prime \nu \mu }\overset{\sim
}{a^{\prime }}_{\mu }=
\label{larmor-finale} \\
&=&\frac{\overset{\frown }{w}^{\prime \sim }}{q}D^{\prime \nu \mu
}\left[ \left( l_{\mu }^{\prime }\cos \phi ^{\prime }+l_{\mu
}^{\prime \prime }\sin
\phi ^{\prime }\right) \right] +\lambda D^{\prime \nu \mu }\left( \overset{%
\sim }{a}_{2}^{\prime }l_{\mu }^{\prime }+\overset{\sim
}{a}_{3}^{\prime }l_{\mu }^{\prime \prime }\right) ,  \notag
\end{eqnarray}%
while the gyrokinetic differential 1-form (which is by
construction gyro-phase independent) reads
\begin{eqnarray}
\overset{\_}{\gamma ^{\prime }(\mathbf{y}^{\prime })} &=&\left( \frac{q}{%
\varepsilon }A_{\mu }^{\prime }+\overset{\_}{u_{\mu }^{\prime
}}\oplus
\lambda q\overset{\_}{a}_{\mu }^{\prime }\right) dr^{\prime \mu }+\frac{1}{2}%
\left\langle (\overset{\sim }{u_{\mu }^{\prime }}\oplus \lambda q\overset{%
\sim }{a}_{\mu }^{\prime })\frac{\partial r_{1}^{\mu }}{\partial
\phi
^{\prime }}\right\rangle d\phi ^{\prime }+  \label{funzionale velcanc} \\
&&+\xi \left[ \left\langle u_{\mu }u^{\mu }\right\rangle -1\right]
ds
\end{eqnarray}%
This expression is formally similar to\ that determined in Ref.1,
except for
the explicit appearance of terms containing the 4-velocity addition law (\ref%
{somma 4vel-1}) on the r.h.s. of Eq.(\ref{funzionale velcanc}),
which have been introduced to preserve their proper physical
interpretation (as 4-velocities) in the corresponding
Euler-Lagrange equations. This implies that non-linear corrections
in the wave-fields are, actually, taken into account in the
gyrokinetic differential 1-form. This result shows the (formal)
intrinsic simplicity of the perturbative approach to GKT here
developed.

\section{Conclusions}

In this paper we have formulated a covariant gyrokinetic approach
for single-particle dynamics utilizing the variational approach
developed in Ref.1. We have shown that:

\begin{itemize}
\item the theory goes beyond the domain of validity of the treatment
previously considered
\cite{Beklemishev1999,Beklemishev2004,Nicolini2005} and includes
the case in which the electric field can locally vanish or result
much smaller than the magnetic field;

\item the treatment includes the effect produced by the presence of
wave-fields (produced by EM and gravitational perturbations). In
particular, we have shown that this generally requires the
introduction of an extended gyrokinetic transformation based on
the introduction of an appropriate 4-velocity perturbative
expansion;

\item a relativistic addition law has been invoked for the 4-velocity
perturbative expansion;

\item non-linear corrections in the wave fields have been take into account
in the perturbative theory to assure that the condition of
physical realizability is satisfied by the transformed
(gyrokinetic) 4-velocity.
\end{itemize}

The simplicity of the approach and its applicability to a wide
range of possible physical situations make the present theory
particularly suitable for applications to the investigation of
relativistic plasmas, potentially both in astrophysics and
laboratory plasmas.

\section*{Acknowledgments}
Work developed in cooperation with the CMFD Team, Consortium for
Magneto-fluid-dynamics (Trieste University, Trieste, Italy). \
Research developed in the framework of the MIUR (Italian Ministry
of University and Research) PRIN Programme: {\it Modelli della
teoria cinetica matematica nello studio dei sistemi complessi
nelle scienze applicate}. The support (A.B) of ICTP (International
Center for Theoretical Physics, Trieste, Italy), (A.B.) University
of Trieste, Italy, (M.T) COST Action P17 (EPM, {\it
Electromagnetic Processing of Materials}) and (M.T. and P.N.) GNFM
(National Group of Mathematical Physics) of INDAM (Italian
National Institute for Advanced Mathematics) is acknowledged.
\section*{Notice}
$^{\S }$ contributed paper at RGD26 (Kyoto, Japan, July 2008).




\newpage
\begin{thebibliography}{99}
\bibitem{Littlejohn} R.G.J. Littlejohn, J. Math. Phys. \textbf{20}, 2445
(1979).

\bibitem{Hahm} T.S. Hahm T. S., W.W. Lee and A. Brizard, Phys. Fluids
\textbf{31,} 1940 (1988).

\bibitem{Mohanti} J. N. Mohanty and K. C. Baral, Phys. Plasmas \textbf{3},
804 (1996) and references therein.

\bibitem{Frank} L. Frank, A. King, D. Raine, \textit{Accretion Power in
Astrophysics}, Cambridge Astrophysics Series:20, 2nd ed.,
Cambridge University Press (1992).

\bibitem{Boghosian} B.M. Boghosian, {\it Covariant Lagrangian methods of
relativistic plasma theory}, PhD. thesis, University of
California, Davis, 1987.

\bibitem{Pozzo} M. Pozzo and M. Tessarotto, Phys. Plasmas \textbf{5,} 2232
(1998).

\bibitem{Beklemishev1999} A. Beklemishev and M. Tessarotto, Phys. Plasmas,
\textbf{6}, 4548 (1999).

\bibitem{Beklemishev2004} A. Beklemishev and M. Tessarotto, A. \& A. \textbf{%
428}, 1 (2004).

\bibitem{Nicolini2005} A. Beklemishev, P. Nicolini and M. Tessarotto,Proc. 24th
RGD, Bari, Italy (July 2004), Ed. M. Capitelli, AIP Conf. Proc.
\textbf{762}, 1283 (2005).

\bibitem{Cremaschini2006} (Ref.1) M. Tessarotto, C. Cremaschini, P. Nicolini
and A. Beklemishev, Proc. 25th RGD (International Symposium on
Rarefied gas Dynamics, St. Petersburg, Russia, July 21-28, 2006),
Ed. M.S. Ivanov and A.K. Rebrov (Novosibirsk Publ. House of the
Siberian Branch of the Russian Academy df Sciences), p.1001
(2007); arXiv:physics/0611114.

\bibitem{Remark} We notice, in fact, that due to the assumed form of the
4-velocity transformation (\ref{II}) the adoption of Lie transform
methods would result cumbersome in the present case. Indeed, Lie
transformations acting on the 4-velocity do not generally satisfy
the relativistic 4-velocity addition law (\ref{somma 4vel-1}).
\end{thebibliography}
\end{document}